\documentclass[aps,pre,preprint,superscriptaddress,nofootinbib]{revtex4-1}
\usepackage{fullpage}
\usepackage{amsmath}
\usepackage{amssymb}
\usepackage{graphicx}
\usepackage{bm}

\usepackage{epsfig}

\usepackage{tikz}
\usepackage{multirow}
\usetikzlibrary{arrows}

\begin{document}

\title{The Relationship Between Local Structure and Relaxation in Out-of-Equilibrium Glassy Systems}
\author{S. S. Schoenholz$^\ddagger$}
\email{schsam@sas.upenn.edu}
\affiliation{Department of Physics, University of Pennsylvania, Philadelphia, Pennsylvania 19104, USA}
\affiliation{Department of Physics and School of Engineering and Applied Sciences, Harvard University, Cambridge, Massachusetts 02138, USA}
\author{E. D. Cubuk$^\ddagger$}
\affiliation{Department of Physics and School of Engineering and Applied Sciences, Harvard University, Cambridge, Massachusetts 02138, USA}
\author{E. Kaxiras}
\affiliation{Department of Physics and School of Engineering and Applied Sciences, Harvard University, Cambridge, Massachusetts 02138, USA}
\author{A. J. Liu}
\email{ajliu@sas.upenn.edu}
\affiliation{Department of Physics, University of Pennsylvania, Philadelphia, Pennsylvania 19104, USA}
\date{\today}
\let\thefootnote\relax\footnotetext{$\ddagger$\hspace{1pc}Equal contribution.}
\begin{abstract}
The dynamical glass transition is typically taken to be the temperature at which a glassy liquid is no longer able to equilibrate on experimental timescales. Consequently, the physical properties of these systems just above or below the dynamical glass transition, such as viscosity, can change by many orders of magnitude over long periods of time following external perturbation. During this progress towards equilibrium, glassy systems exhibit a history dependence that has complicated their study. In previous work, we bridged the gap between structure and dynamics in glassy liquids above their dynamical glass transition temperatures by introducing a scalar field called ``softness'', a quantity obtained using machine learning methods. Softness is designed to capture the hidden patterns in relative particle positions that correlate strongly with dynamical rearrangements of particle positions. Here we show that the out-of-equilibrium behavior of  a model glassforming system can be understood in terms of softness. To do this we first demonstrate that the evolution of behavior following a temperature quench is a primarily structural phenomenon: the structure changes considerably, but the relationship between structure and dynamics remains invariant. We then show that the history-dependent relaxation time can be robustly computed from structure as quantified by softness. Together, these results motivate the use of softness to characterize the history dependence of glasses.
\end{abstract}

\maketitle

In liquids cooled quickly enough so that crystallization is avoided, the dynamics become increasingly sluggish~\cite{angell95,sastry98,debenedetti01} until $\tau_{eq}$, the time required for the system to equilibrate, exceeds experimentally-accessible time scales at what is called the dynamical glass transition temperature, $T_g$.  At time scales shorter than $\tau_{eq}$, quantities such as the potential energy, pressure, etc. and time correlation and response functions of these quantities depend not only on state variables such as the temperature $T$ and density $\rho$, but also on the history of the system as described by its path through $(T,\rho)$-space~\cite{ediger96,angell2000,berthier13}.  This is the situation following a rapid temperature quench from the liquid into a supercooled state. As the system evolves at a fixed temperature following the quench, the dynamics slow down, the average energy barrier height increases, and energy and pressure decrease~\cite{kob97_2,kob00,rottler05,warren10}. This process continues indefinitely if the final temperature is below $T_g$ but stops once the system has reached equilibrium if the temperature is above $T_g$.  

In recent papers, we introduced a machine learning approach to construct a ``softness'' field, $S$~\cite{rottler14,schoenholz14,cubuk15,schoenholz16,cubuk16}, using local structural descriptors~\cite{behler07,cubuk14}.  The softness of a particle quantifies its local structural environment and is designed to correlate strongly with its dynamics; the higher the particle's softness, the more likely it is to rearrange.  Softness can therefore be viewed as a structural order parameter for the dynamics.  In particular, we demonstrated~\cite{schoenholz16} that the probability for a particle of softness $S$ to rearrange obeys an Arrhenius dependence with temperature, given by $P_R(S) = \exp[\Sigma(S)-\Delta E(S)/kT]$ where $\Delta E(S) \approx e_1 - e_2S$ and $\Sigma(S) \approx z_1 - z_2S$.  Thus, particles of softness $S$ must confront energy barriers of order $\Delta E(S)$ in order to rearrange.  For convenience, $P_R(S)$ may be rewritten as
\begin{equation}
P_R(S) = \exp\left[(e_1-e_2S)\left(\frac 1{T_0} - \frac 1T\right)\right]. \label{eq:prarrh}
\end{equation}
with $T_0 = e_2/z_2 \approx e_1/z_1$ being the onset temperature for glassy dynamics.  The advantage of softness is that it simplifies the description of the dynamics of glassy liquids significantly. The relaxation is non-exponential in time because particles with different softnesses relax exponentially with different rates, while the dynamics are heterogeneous because particles with different softness rearrange with different probabilities~\cite{schoenholz16,cubuk16}.  

Here we show that a framework built on softness provides a coherent description of the out-of-equilibrium behavior of glassy liquids both above and below $T_g$. As these systems evolve in time, their softnesses change dramatically.  For particles of a given softness, however,  we find that the probability of a rearrangement, $P_R(S)$, \emph{remains unchanged}.  Even deep inside the glass state, $P_R(S)$ retains its simple Arrhenius form [Eq.~(\ref{eq:prarrh})] with the identical prefactors that we identified in the supercooled liquid.  Thus, the characteristic multiplicity $\Sigma(S)$ and energy barrier $\Delta E(S)$ are independent of the age of the glass. This surprising result implies that the changing behavior of glassy systems as they age or approach equilibrium is primarily structural in origin;  the structure, as quantified by softness, changes, but the relationship between structure and dynamics remains invariant. We exploit this realization to show that the relaxation time of these systems, both in and out of equilibrium, can be predicted accurately from a simple ``mean-field'' model of relaxation. This mean-field model is fundamentally incompatible with several long-standing semi-empirical equations for the relaxation time, including the Vogel-Fulcher-Tammann (VFT) equation. The structural nature of equilibration along with the success of our mean-field model suggests that protocol dependence of out-of-equilibrium glassy liquids (as quantified by the knowledge of the path through $(T,\rho)$-space) can be replaced by the instantaneous softness distribution. This realization should make the study of nonequilibrium glassy liquids substantially more tractable.

We first summarize our method for computing the softness; for a more detailed description, see Refs.~\cite{cubuk15,schoenholz16,cubuk16}.  We characterize the local structural environment of a central particle $i$ in terms of a set of $M=166$ structure functions, $G_\alpha(i)$~\cite{schoenholz16}.  The structure of the particle is represented as a point in $M$-dimension space, $\mathbb R^M$, where each axis corresponds to a different structure function. We then select a ``training set'' of particles and calculate their associated structure functions.  Half of the particles are chosen to be those that are about to rearrange, while the other half have not rearranged for a long time.  We use the method of Support Vector Machines (SVMs)~\cite{SVM,libsvm} to find the hyperplane that best separates these two groups of particles in $\mathbb R^M$. Once the hyperplane is identified, the structure around any particle can be characterized by computing the signed distance of its position in $\mathbb R^M$ to the hyperplane.  This quantity is the ``softness''. Throughout this paper we use a hyperplane constructed for the system at temperature $T = 0.47$, which is above its dynamical mode-coupling temperature at $T_{MC} \approx 0.435$.

To study equilibration and aging, we follow the procedure outlined by Kob and Barrat~\cite{kob97_2} using molecular dynamics.  We first equilibrate an 80:20 binary Lennard-Jones mixture of $N=10,000$ particles at a high temperature, $T_I$. After equilibration the system is instantaneously quenched to a temperature $T_F$. If $T_F>T_g$ then the system will reach equilibrium at some measurable time. If $T_F<T_g$ then the behavior of the system will continue to evolve as the system ages on timescales accessible to our simulations. We track various properties of the system as a function of the waiting time, $t_w$, following the quench to $T_F$. At exponentially-spaced time intervals we take snapshots of the system.  We then quench each snapshot to its inherent structure using a combination of conjugate gradient minimization and FIRE minimization~\cite{bitzek06} and calculate the softness for the inherent structure. In this work we always choose the central particle to be of species A (large), but our results also hold for species B.

\begin{figure}[h!]
\centering
\includegraphics[width=0.9\textwidth]{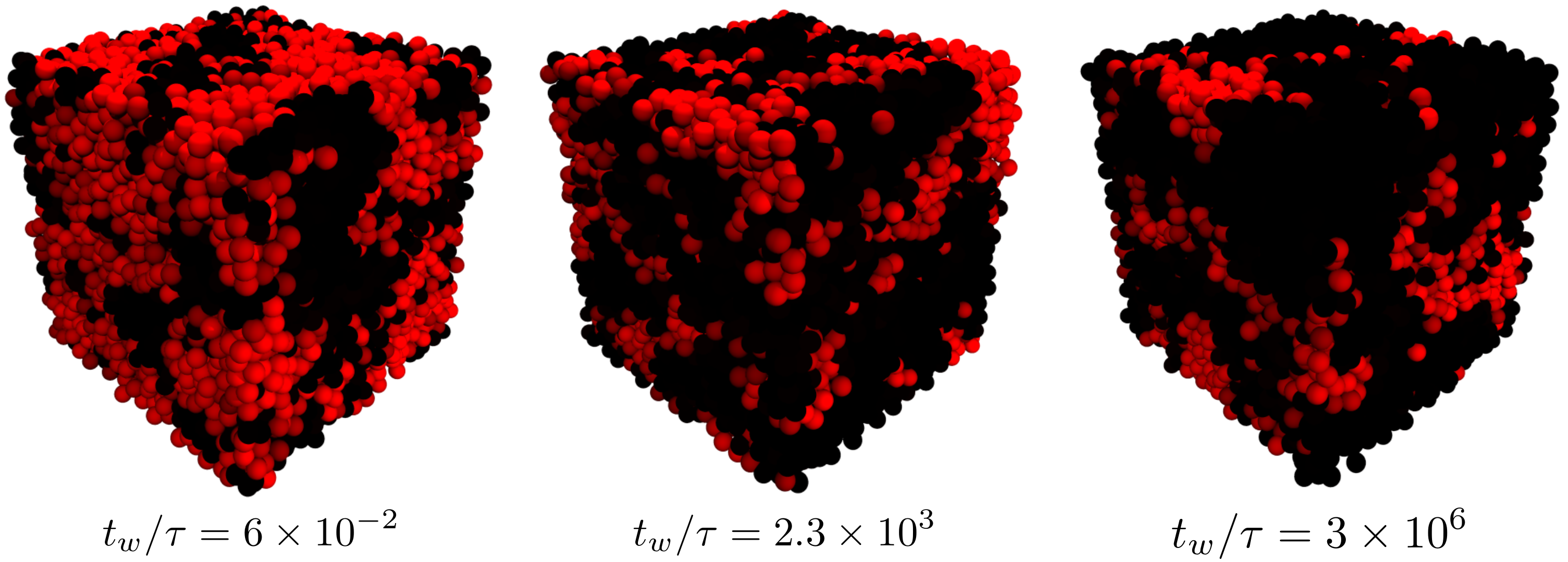}
\caption{Three snapshots of an aging glass at three different waiting times, $t_w/\tau$ prepared with $T_I = 1.0$ and $T_F=0.4$. Particles with $S>1$ are red while particles with $S<1$ are black.}  
\label{fig:system}
\end{figure}

In Fig.~\ref{fig:system} we show inherent structures of the system at three different values of $t_w$ that vary by eight orders of magnitude, with each particle colored according to its softness. We see a stark change as the system is aged from early times, when it is predominantly soft (red), to later times, when much of the system has hardened. Since the softness characterizes the local structure around each particle, these images already demonstrate qualitatively that significant structural changes occur in glasses during aging.

For a quantitative analysis of evolution during aging, we investigate the connection between the changing structure and the increasingly sluggish dynamics of the aging system. The slow dynamics in the glass can be quantified in terms of the relaxation time $\tau_\alpha(t_w)$ at a time $t_w$ following the quench from $T_I$ to $T_F$.   The relaxation time is defined in terms of the self intermediate scattering function,
\begin{equation}
F_s(\bm q_{\text{max}},t;t_w) = \sum_j e^{-i\bm q_{\text{max}}\cdot [\bm r_j(t+t_w) - \bm r_j(t_w)]}
\end{equation}
where $\bm q_{\text{max}} = 7.05$ is chosen to be wave vector at the first peak of the static structure factor. We take the relaxation time, $\tau_\alpha(t_w)$, to be the time at which $F_s(\bm q_{\text{max}},t;t_w)$ decays to $e^{-1}$. 

\begin{figure}[h!]
\centering
\includegraphics[width=0.9\textwidth]{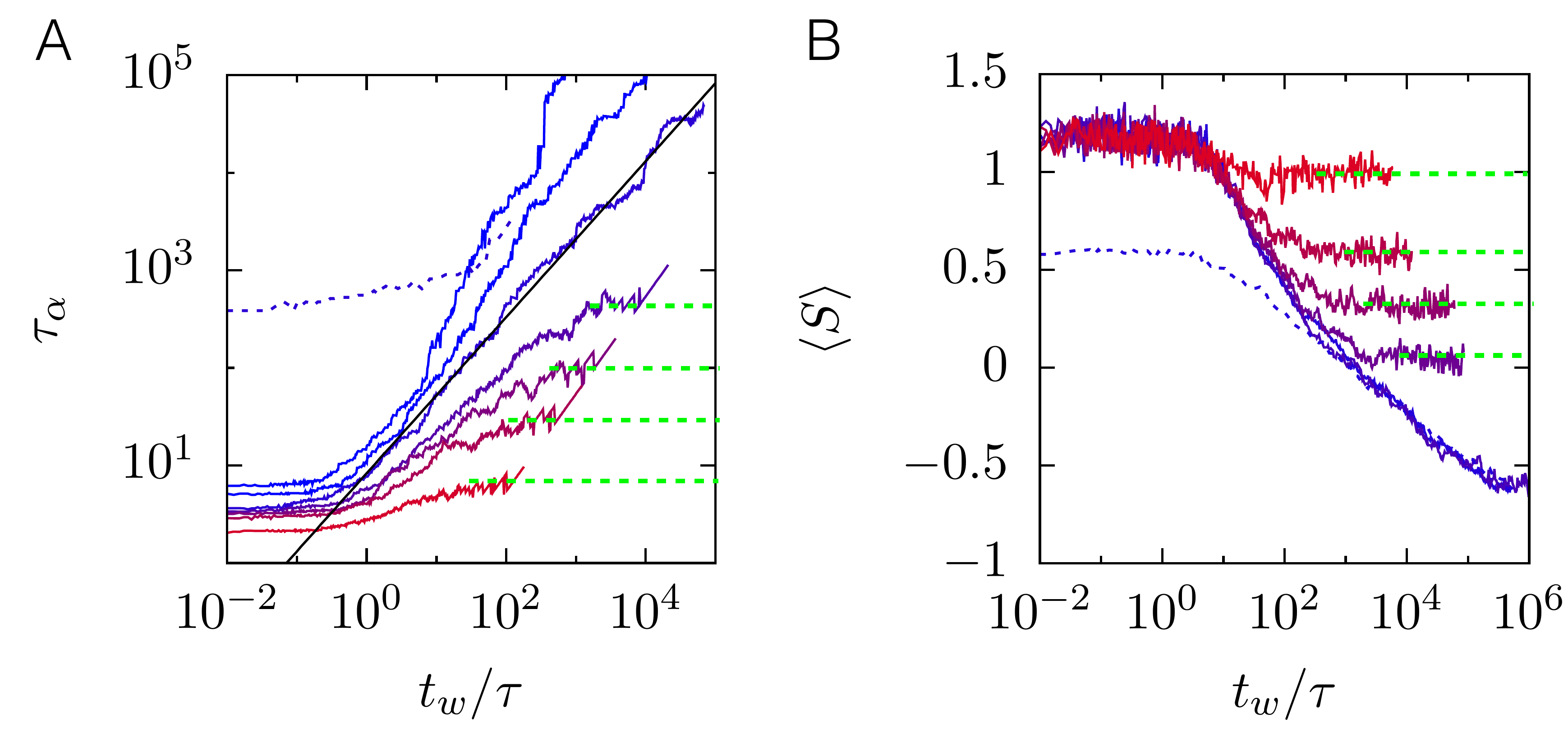}
\caption{The aging of glasses and out-of-equilibrium supercooled liquids. The dotted blue line represents the system with $T_I = 0.8$, whereas all other systems have the $T_I = 1.0$. Dashed green lines indicate equilibrium values in both plots. (A) The relaxation time $\tau_\alpha$ as a function of $t_w/\tau$. (B) The average softness as a function of $t_w/\tau$. Before equilibration all protocols follow the same ``universal'' curve.}  
\label{fig:softness}
\end{figure}

We plot $\tau_\alpha(t_w)$ in Fig.~\ref{fig:softness} (a) for $t_w$ varying over several orders of magnitude for several different choices of $T_I$ and $T_F$. For all choices studied, we find that at very short times, $t\lesssim\tau_x$, the relaxation time is approximately constant,  where $\tau_x$ is the timescale of the crossover from ballistic to caged dynamics. As expected, the initial relaxation time at $t_w \lesssim \tau_x$ is significantly longer for colder initial temperatures.  

At later times the increase of relaxation time is well-described by a power law, $\tau_\alpha\sim t_w^\beta$. With $T_I = 1.0$ and $T_F = 0.4$ we find $\beta\sim 0.8$, in agreement with Ref.~\cite{kob97_2}. As expected for $T_F < T_g$, $\tau_\alpha$ increases without bound from its initial value when $t_w \approx \tau_x$ at short times to $10^5\tau$ for our longest aged sample. By contrast, when $T_F > T_g$, $\tau_\alpha$ increases until the sample equilibrates at some finite $t_w$. At this point, the relaxation time flattens out at its equilibrium value. 

In Fig.~\ref{fig:softness}(B) we plot the mean softness of particles, $\langle S(t_w)\rangle$ as a function of waiting time $t_w$. The initial average softness depends on $T_I$ and -- like the relaxation time -- is constant for $t\lesssim \tau_x$.  At later times,
as already suggested qualitatively by Fig.~\ref{fig:system}, the average softness decreases significantly as the system ages for all values of $T_I$ and $T_F$ studied. While the mean softness changes, the distribution of softness remains approximately Gaussian throughout the aging process with constant variance (see supplementary information.) 
Remarkably, for $t>\tau_x$ the average softness appears to decrease as an approximately logarithmic function of $t_w$ that depends neither on $T_I$ nor on $T_F$. For $T_F>T_g$, the mean softness levels out at its equilibrium value when the system reaches equilibrium. For $T_F<T_g$, the mean softness decreases with no sign of a plateau, as expected.

\begin{figure}[h!]
\centering
\includegraphics[width=0.9\textwidth]{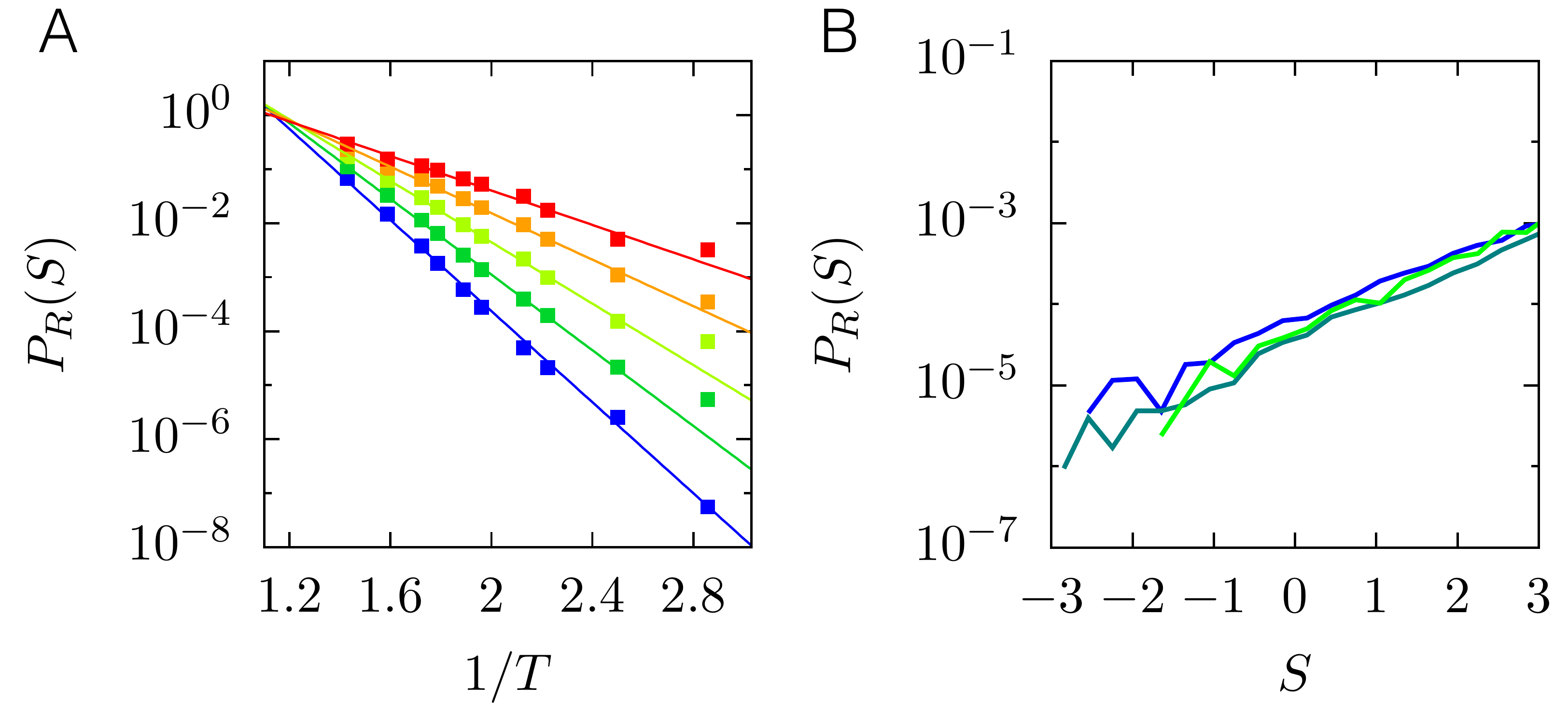}
\caption{The relationship between structure and dynamics. (a) The probability of rearrangement as a function of $1/T$ for particles of different softness from $S=-3$ (blue) to $S=3$ (red). The Arrhenius result holds deep into the glassy regime. (b) The probability of rearrangement as a function of softness, $S$ at $T_F=0.40$ for three different waiting times from $t_w/\tau = 3\times10^3 $ (blue) to $t_w/\tau= 5\times10^6 $ (green). There is no statistically significant change in the probability of rearrangement as a function of age.}  
\label{fig:rearrange}
\end{figure}

To connect structure (softness) with dynamics, we consider the softness-dependent probability of rearrangement, $P_R(S)$~\cite{schoenholz16}.  We calculate $P_R(S)$, the fraction of particles of softness $S$ that are rearranging, as a function of waiting time $t_w$ following quenches from $T_I$ to $T_F$. In Fig.~\ref{fig:rearrange}(A) we plot $P_R(S)$ as a function of $1/T$ from high temperatures, where our system behaves nearly as a simple liquid, down to $T=0.35$, which is the lowest temperature for which we were able to obtain enough statistics to reliably compute $P_R(S)$.  Note that $T=0.35$ is well below the dynamical glass transition temperature for this system. Fig.~\ref{fig:rearrange}(A) also shows the Arrhenius fits that we obtained in Ref.~\cite{schoenholz16} for the temperature range $0.47 < T< 0.70$. The excellent agreement between the fits and the data from $T < 0.47$ shows that the probability of rearrangement continues to have Arrhenius form well below the glass transition temperature. In Fig.~\ref{fig:rearrange}(B) we plot $P_R(S)$ as a function of softness at $T=0.40$ for three different waiting times, $t_w$, that vary by three orders of magnitude. Surprisingly, $P_R(S)$ is approximately independent of age.  

The results of Fig.~\ref{fig:softness}(B) and Fig.~\ref{fig:rearrange}(A) imply that the description of the aging process is simplified considerably when viewed through the lens of softness.  As a glass ages, it has long been recognized that the average energy barrier increases as the system becomes trapped in deeper and deeper minima~\cite{kob00,rottler05}.  Our results show that for particles of a given softness, the energy barrier is unchanged.  The average energy barrier increases with age simply because the distribution of softness shifts to lower values.  Thus, the increasing relaxation time of glasses and supercooled liquids during aging is primarily structural in origin.  Our results also imply that the history-dependent behavior of glasses can be understood in terms of local structure as quantified by the softness field.

We now consider the relationship between the relaxation time $\tau_\alpha$ and softness. Earlier~\cite{schoenholz16}, we showed that the time-dependence of relaxation could be predicted from softness by combining the probability of rearrangement, $P_R(S)$, with a ``softness propagator'', $G(S,S_0,t)$, which measures the probability that a particle with softness $S_0$ at $t=0$ will have a softness $S$ at a time $t$. The softness propagator accounts for changes in the softness of a particle due to nearby rearrangements even when the particle itself does not rearrange. In practice we found that $G(S,S_0,t)$ is similar to the Green's function for a directed diffusion process in which particles that begin with a softness $S_0$ evolve towards softnesses closer to $\langle S\rangle$ with time.  This suggests that the scaling of the relaxation time is controlled by the average softness as follows,
\begin{equation}\tau_\alpha\sim\frac1{P_R(\langle S\rangle)} \sim \exp\left[(\alpha_1\langle S\rangle-\alpha_2)\left(\frac1{T_0} - \frac1T\right)\right]
\label{eq:tauS}
\end{equation} 
where $\alpha_1$ and $\alpha_2$ are free temperature-independent parameters that arise because we are measuring relaxation with the intermediate scattering function instead of the overlap function.

\begin{figure}[ht!]
\centering
\includegraphics[width=0.9\textwidth]{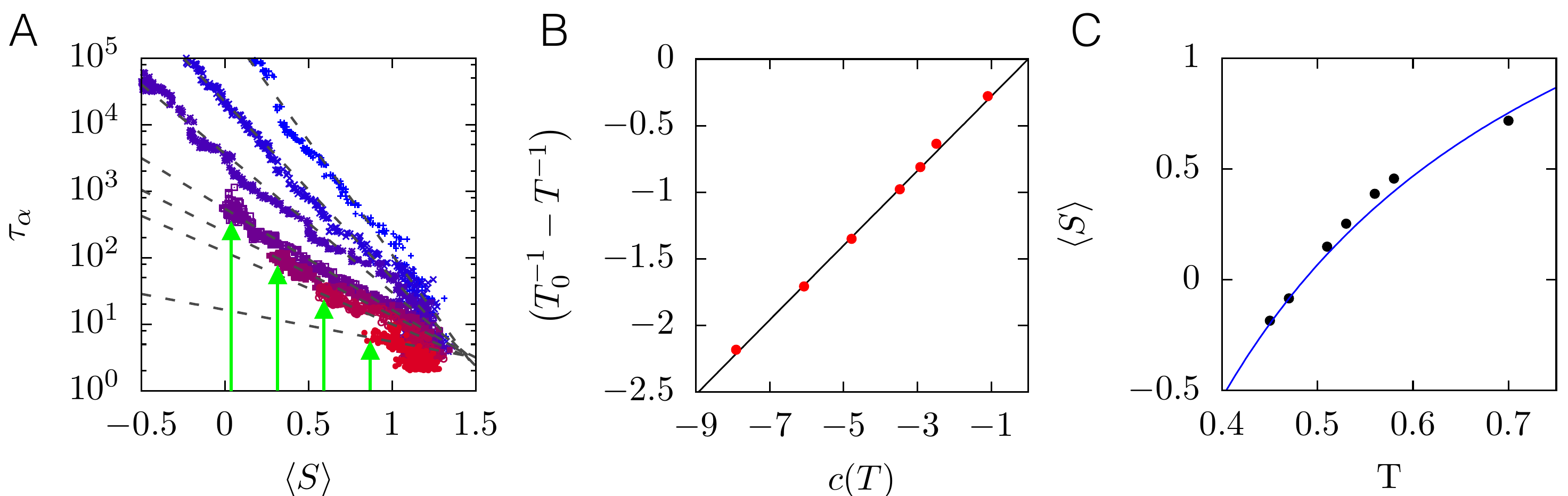}
\caption{The relationship between relaxation time and structure. (A) The relaxation time, $\tau_\alpha$, as a function of average softness. Dashed lines are best fits to the prefactor of the exponential dependence $\tau_\alpha\sim\exp(c(T)\langle S\rangle)$. Green arrows indicate the equilibrium softness and relaxation time for supercooled liquids. (B) Comparison of $c(T)$ with the predicted dependence of $c(T)\sim T_0^{-1} - T^{-1}$, showing agreement between the measured and theoretical dependence. (C) A comparison of the average equilibrium softness with predictions from a parabolic form for relaxation proposed in Ref.~\cite{elmatad09}. }  
\label{fig:relaxation_softness}
\end{figure}

To test this prediction we plot, in Fig.~\ref{fig:relaxation_softness}(A), the relaxation time $\tau_\alpha$ as a function of the mean softness, $\langle S\rangle$. Green arrows denote the equilibrium value of average softness for systems above the glass transition. Indeed, we find that the relaxation time depends on the average softness exponentially, as predicted by Eq.~(\ref{eq:tauS}). For each final temperature $T_F$, we fit $\log\tau_\alpha$ \emph{vs.} $\langle S\rangle$ to straight lines (grey dashed line fits in Fig.~\ref{fig:relaxation_softness}(A)) and denote the slope by $c(T)$. Finally, we plot in Fig.~\ref{fig:relaxation_softness}(B) $c(T)$ against our prediction in Eq.~(\ref{eq:tauS}). We find excellent agreement at all temperatures with $\alpha_1\approx 3.6$ (note that $\alpha_2$ is a free overall multiplicative constant that does not affect this measurement).  This agreement holds both in equilibrium glassy liquids and in aging systems below $T_g$. This provides strong evidence that Eq.~(\ref{eq:tauS}) is a robust descriptor of relaxation in glassy systems both in and out of equilibrium. 

It is interesting to compare the form of relaxation found in Eq.~(\ref{eq:tauS}) with previous models of glassy relaxation. In particular, we consider a parabolic form~\cite{elmatad09,keys11}, the Vogel-Fulcher-Tammann (VFT) form, and the B\"assler law~\cite{berthier11_3} given respectively by,
\begin{align}
    \tau_\alpha^C &= \exp\left[J\left(\frac1{T_0}-\frac1T\right)^2\right]\label{eq:tauModelChandler}\\
    \tau_\alpha^{VFT} &= \exp\left[A\frac1{T-T_{VFT}}\right]\\
    \tau_\alpha^{B} &= \exp\left[B\left(\frac{T_B}T\right)^2\right]
    \label{eq:tauModelBassler}
\end{align}
where we allow $J, T_{VFT}, T_B, A,$ and $B$ to be free parameters to account for differences in protocol. Each of these laws has been used to fit a large set of experimental relaxation time data for glassy liquids over many decades of relaxation time. To make the comparison we consider the temperature dependence of $\langle S\rangle$ implied by combining Eq.~(\ref{eq:tauS}) with each of the models Eq.~(\ref{eq:tauModelChandler})-(\ref{eq:tauModelBassler}). We then compare the implied $\langle S\rangle$ for equilibrium supercooled liquids with our expectations. The three different functional forms for $\tau_\alpha$ give, respectively, 
\begin{align}
    \langle S\rangle^C &= \frac{\alpha_2}{\alpha_1} + \frac{J}{\alpha_1}\left(\frac1{T_0}-\frac1T\right)\\
    \langle S\rangle^{VFT} &= \frac{\alpha_2}{\alpha_1} + \frac{A}{\alpha_1(T-T_{VFT})(T_0^{-1}-T^{-1})}\\
    \langle S\rangle^{B} &= \frac{\alpha_2}{\alpha_1} + \frac B{\alpha_1}\left(\frac{T_B}T\right)^2\frac1{T_0^{-1}-T^{-1}}.
\end{align}
Both $\langle S\rangle^{VFT}$ and $\langle S\rangle^{B}$ feature divergences including one at $T_0$. We can easily measure $\langle S\rangle$ at $T_0$; we observe a finite mean softness there.  Indeed, $\langle S\rangle$ can be represented as a sum over local density~\cite{cubuk16} so it cannot diverge. We therefore conclude that Eq.~(\ref{eq:tauS}) is inconsistent with the VFT equation and the B\"assler law for relaxation. This argument does not preclude modified forms of these laws that are asymptotically the same at lower temperatures with no pole at $T_0$.  The VFT form is more problematic since it also predicts a divergence of $\langle S \rangle$ at $T_0$, implying that the functional dependence of Eq.~\ref{eq:tauS} on $\langle S \rangle$ would need to change at lower temperatures or higher waiting times in order for relaxation time to diverge at a nonzero value of $T_0$. We plot in Fig.~\ref{fig:relaxation_softness}(C) a comparison of $\langle S\rangle$ with the prediction from the parabolic form, which shows strong agreement between the model and our measured valued for the mean softness. 


\begin{figure}[h!]
\centering
\includegraphics[width=0.65\textwidth]{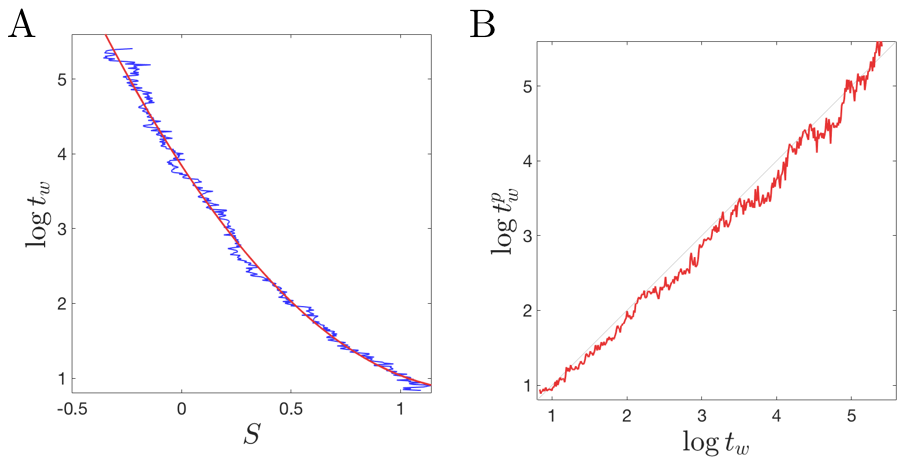}
\caption{Prediction of the age of a glass from its structure. (A) $t_w$ \emph{vs.} $S$ for the system at $T_F = 0.40$ (blue), and a parabollic fit (red). (B) Predicted $t_w$ \emph{vs.} $t_w$ of the system at $T_F = 0.35$. Predictions are based on the parabolic fit from (A).}  
\label{fig:age_prediction}
\end{figure}

Finally, we investigate the question of whether or not the age of a glass may be inferred from its structure alone. To this end, we utilize the observation that the average softness of a system seems to follow the same function of $t_w$, independent of $T_I$ and $T_F$. Thus, we start by fitting $t_w$ of the system at $T_F=0.40$ by a smooth curve, as a function of its average softness. A quadratic polynomial of average softness is sufficient to predict $t_w$ of this system (Fig.~\ref{fig:age_prediction}(A)). We test the accuracy of this fit by predicting $t_w$ of the system at $T_F=0.35$, using only its average softness value: $t_w$ can be accurately predicted, even for systems at different temperatures, only by using the average softness of the system (Fig.~\ref{fig:age_prediction}(B)). We suggest that this approach can be used to date disordered materials of unknown age, as long as a model can be fit to the age of another system of the same material as a function of its average softness or other sufficiently descriptive structural quantities that can be measured at different ages. It would be interesting to study systems that are aged longer in experiments or by parallelizing molecular dynamics in time scale~\cite{perez15}.    

Our results show that the concept of softness is useful even for systems out of equilibrium at temperatures below the dynamical glass transition. Indeed, it would appear that history-dependent behavior in glasses can be understood in terms of local structure as quantified by the softness field, and that the connection between softness and the relaxation time is remarkably simple and independent of age.  A common critique of numerical results such as the ones presented here in glass transition studies is that the timescales accessible in simulation are short compared to those observable in experiments.  As a result, studies restricted to the equilibrium behavior of glassy liquids necessarily probe only properties at relatively high temperatures.  It is encouraging that we observe exactly the same functional form in the equilibrium liquid and well inside the aging glass state for: (1) the relation of softness to the probability of a rearrangement, and (2) the relation of the relaxation time to average softness.  This agreement suggests that our results for these relations are not hampered by limitations of computational modeling. These results, together with our demonstration that aging is structural, provide evidence that history dependence in glasses can be quantified using softness.

\bibliography{bibliography}

\end{document}